\documentclass[sigconf,screen]{acmart}

\setcopyright{none}
\acmConference{}{}{}
\acmYear{}
\copyrightyear{}
\acmISBN{}
\acmDOI{}

\renewcommand{\footnotetextcopyrightpermission}[1]{}
\AtBeginDocument{\fancyhead[LE,RO]{}}

\usepackage{amsmath}
\usepackage{tabularx,array}
\usepackage[caption=false]{subfig}
\usepackage{multirow}
\usepackage{adjustbox}
\usepackage{pifont}
\providecommand{\taxyes}{\ding{51}}
\providecommand{\taxno}{\ding{55}}

\newcommand{\simname}{GIPSim}
\newcommand{\uop}{\textmu{}op}
\newcolumntype{Y}{>{\raggedright\arraybackslash}X}
\keywords{architectural simulation, trace replay, out-of-order execution,
parallel simulation, synchronization}

\begin{document}

\title{Scaling Architectural Simulation to Giga Instructions Per Second}

\author{Zhouyi Shen}
\affiliation{%
  \institution{University of Science and Technology of China}
  \city{Hefei}
  \country{China}
}
\email{squirrel@mail.ustc.edu.cn}

\author{Hao Tang}
\affiliation{%
  \institution{University of Science and Technology of China}
  \city{Hefei}
  \country{China}
}
\email{th20030818@mail.ustc.edu.cn}

\author{Ruike Liu}
\affiliation{%
  \institution{University of Science and Technology of China}
  \city{Hefei}
  \country{China}
}
\email{liuruike@mail.ustc.edu.cn}

\author{Jingwei Sun}
\affiliation{%
  \institution{University of Science and Technology of China}
  \city{Hefei}
  \country{China}
}
\email{sunjw@ustc.edu.cn}

\author{Guangzhong Sun}
\affiliation{%
  \institution{University of Science and Technology of China}
  \city{Hefei}
  \country{China}
}
\email{gzsun@ustc.edu.cn}

\renewcommand{\shortauthors}{Shen et al.}

\begin{abstract}

The cost of detailed architectural simulation constrains design-space exploration, while parallelization across simulated cores offers limited host concurrency for single-core and small multicore targets. We present GIPSim, an instruction-driven simulator that derives instruction timing from dependencies and resource constraints, reducing the overhead of updating pipeline state every simulated cycle. It applies synchronization-guided temporal decomposition to recorded execution, exposing host parallelism beyond the target core count.
\simname{} replays the resulting segments in parallel and aggregates their durations through a dependency graph to estimate overall execution time, with replay contexts reconstructed using retained microarchitectural state, warmup, and synchronization-guided cache eviction.
This accounts for synchronization waits without requiring host workers to follow target execution order. We evaluate GIPSim using QEMU traces of application and operating system execution on multi-core x86-64, AArch64, and RV64GC systems.
\simname{} achieves an average single-node simulation throughput of 13.8 million instructions per second (MIPS) with one worker and 1{,}010 MIPS with 96 workers, peaking at 1{,}421 MIPS. The corresponding mean absolute percentage errors in execution cycles relative to gem5 are 7.52\% and 7.43\%, respectively.

\end{abstract}

\maketitle

\section{Introduction}
\label{sec:introduction}

Architectural simulation provides a quantitative basis for evaluating processor designs before implementation. 
The widely used gem5~\cite{gem5} simulator combines cycle-driven detailed CPU models with a discrete-event framework, incurring substantial overhead from repeated state updates and event processing. 
Published measurements~\cite{partigem5} report gem5 timing-mode throughput of 0.01--0.1 million instructions per second (MIPS), approximately five to six orders of magnitude below native hardware execution rates. This poor simulation performance delays feedback on individual design choices and restricts the number of workload--configuration pairs that can be evaluated, limiting the scale of architectural design-space exploration.

To alleviate this bottleneck, prior work pursues three complementary directions: improving timing efficiency~\cite{interval,zsim,prophet}, distributing simulation work across host workers~\cite{graphite,sniper,partigem5}, and sampling execution regions to reduce the amount of detailed simulation~\cite{simpoint,smarts,looppoint}. This paper focuses on timing efficiency and parallel simulation. Sampling is complementary to these efforts: it determines which regions to simulate in detail, while timing models and parallel simulation determine how those regions are evaluated.

Instruction-driven simulation improves timing efficiency by deriving instruction timing from dependencies and resource constraints rather than repeatedly evaluating pipeline state each cycle~\cite{zsim,prophet}. It can model out-of-order execution and resource contention with less state-maintenance overhead, but its accuracy depends on how faithfully scheduling and memory interactions are represented. Simplified issue ordering can change how instructions compete for resources, while incomplete translation and memory timing can distort stalls and the overlap between computation and memory accesses.

Parallel simulation complements improvements in timing efficiency by distributing modeling work across host workers.
\emph{Spatial decomposition} assigns target cores or components to different workers, preserving each component's state as execution progresses. Graphite, Sniper, and ZSim exploit core-level parallelism~\cite{graphite,sniper,zsim}, while parti-gem5 parallelizes detailed gem5 CPU and memory models~\cite{partigem5}. 
Shared-resource interactions require workers to coordinate progress in simulated time. Core-level partitioning also limits concurrency to the target core count, leaving much of a large host underused when simulating single-core or small multicore targets. Running multiple simulation instances concurrently improves aggregate throughput, but does not shorten an individual evaluation.

\emph{Temporal decomposition} exposes parallel work along the execution timeline. Barrier-Interval Time-Parallelism~\cite{barrier-interval} and ChunkedTejas~\cite{chunkedtejas} construct coarse-grained intervals spanning participating cores and jointly simulate the components within each interval. Independently replaying per-core trace segments would expose parallel work across both execution positions and target cores, while introducing two linked challenges.
First, the decomposition must identify suitable boundaries and represent the relative positions of segments, enabling their durations to be properly combined to predict overall execution time. 
Second, independent replay requires a suitable simulation context for each segment, including an approximation of its entry state and estimates of resource contention from other cores during execution.

\begin{table}[t]
    \centering
    \caption{Representative approaches to simulation work decomposition and timing organization. MC: multicore target support; SD/TD: spatial/temporal decomposition.}
    \label{tab:taxonomy}
    \small
    \setlength{\tabcolsep}{2pt}
    \renewcommand{\arraystretch}{1.15}
    \renewcommand{\tabularxcolumn}[1]{m{#1}}
    \begin{tabularx}{\columnwidth}{
        >{\raggedright\arraybackslash}m{0.33\columnwidth}
        *{3}{>{\centering\arraybackslash}m{0.06\columnwidth}}
        >{\raggedright\arraybackslash}X
    }
        \toprule
        \textbf{Related Work}
        & \textbf{MC}
        & \textbf{SD}
        & \textbf{TD}
        & \textbf{Timing Organization} \\
        \midrule
        gem5~\cite{gem5}
        & \taxyes & \taxno & \taxno
        & Sequential event processing \\
        \midrule
        Graphite~\cite{graphite}; Sniper~\cite{sniper}; ZSim~\cite{zsim}; Prophet~\cite{prophet}; parti-gem5~\cite{partigem5}; SplitSim~\cite{splitsim}
        & \taxyes & \taxyes & \taxno
        & Coordinated component progression \\
        \midrule
        LEGOSim~\cite{legosim}
        & \taxyes & \taxyes & \taxno
        & Staged component replay \\
        \midrule
        Nguyen et al.~\cite{traceparallel1997}; DiST~\cite{girbal2003dist}
        & \taxno & \taxno & \taxyes
        & Independent interval replay + timing aggregation \\
        \midrule
        Pandey et al.~\cite{pandey2022gpu}
        & \taxno & \taxno & \taxyes
        & Independent sub-trace replay + correction \\
        \midrule
        Barrier-Interval Time-Parallelism~\cite{barrier-interval}
        & \taxyes & \taxno & \taxyes
        & Joint multicore simulation within intervals \\
        \midrule
        ChunkedTejas~\cite{chunkedtejas}
        & \taxyes & \taxyes & \taxyes
        & Joint multicore simulation within intervals \\
        \midrule
        \textbf{\simname{}}
        & \taxyes & \taxyes & \taxyes
        & \textbf{Independent per-core replay + DAG-based time reconstruction} \\
        \bottomrule
    \end{tabularx}
\end{table}

We present \simname{}, a trace-driven performance simulator that combines efficient instruction-driven timing with \emph{synchronization-guided temporal decomposition} for fine-grained independent replay.
For efficient and accurate instruction timing, its engine explicitly schedules ready operations out of order under resource constraints and uses timestamps to couple instruction progress with address translation and memory service. This captures instruction scheduling and memory-induced stalls without repeated cycle-by-cycle pipeline updates.

\simname{} enables independent parallel replay by decoupling host replay order from target execution order while preserving execution dependencies in timing prediction. Two complementary designs support this decoupling:
(1) To reconstruct overall execution time, \simname{} partitions each core's trace at synchronization points and represents the segments and their dependencies in a directed acyclic graph (DAG). Critical-path analysis aggregates segment durations, accounting for synchronization waits.
(2) To provide replay contexts, \simname{} approximates entry states using retained microarchitectural state and history-based warmup. Synchronization-guided cache eviction additionally approximates the cache effects of cross-core writes, while an analytical model estimates additional main-memory contention among active cores. Workers can thus replay independently without predecessor timing results or continuous cross-component coordination. Independent replay of core-local segments also provides spatial decomposition, as summarized in Table~\ref{tab:taxonomy}.

We evaluate \simname{} using QEMU-generated traces from SPEC CPU, PARSEC, and the NAS Parallel Benchmarks (NPB). The target systems include one- and eight-core x86, ARM, and RISC-V configurations. Across 48 workload--hardware pairs, average throughput on a single host node increases from 13.8 MIPS with one worker to 1{,}010 MIPS with 96 workers. Single-core targets achieve 68.1--76.0$\times$ speedup, directly demonstrating parallelism beyond the target core count. The mean absolute percentage error (MAPE) in predicted cycles relative to gem5 is 7.52\% with one worker and 7.43\% with 96 workers. Throughput measurements include initialization, replay and post-processing but exclude trace collection.

This paper makes three contributions:
\begin{itemize}
\item An instruction-driven timing model that explicitly schedules operations out of order and couples core execution with address translation and memory service through timestamps, preserving key timing interactions without repeated cycle-by-cycle pipeline updates.

\item A synchronization-guided temporal decomposition for independent parallel replay and dependency-based execution-time reconstruction, exposing host concurrency beyond the target core count.

\item A multi-ISA implementation and evaluation quantifying how replay parallelism and state reconstruction affect prediction accuracy and throughput across workloads and processor configurations.
\end{itemize}

\section {Related Work}
\label{sec:related}

\subsection{Architectural Simulation}
\label{sec:simulation-basics}

Architectural simulation models processor and memory behavior for design exploration before implementation. \emph{Functional simulators} reproduce instruction effects on architectural state. QEMU~\cite{qemu} uses dynamic binary translation for near-native functional simulation and serves as our trace frontend. \emph{Performance simulators} additionally model microarchitectural timing to predict execution time. The widely adopted gem5~\cite{gem5,gem5-20} provides detailed processor and memory models. Its capabilities and prediction accuracy have been extensively studied and validated~\cite{akram2019survey,shao2016co,huppert2021memory}, making it our accuracy reference.

In \emph{execution-driven} simulation~\cite{zsim,mcqemu,gem5}, functional execution and timing proceed together, allowing timing to affect execution progress and interleaving. In \emph{trace-driven} simulation~\cite{rico2011trace,elastic}, a separate timing backend follows recorded execution without feedback to the functional frontend. \simname{} adopts this design because previously generated traces simplify the distribution of replay work across host workers and support multiple timing evaluations without repeating functional execution.

Sampling reduces the work requiring detailed simulation. BarrierPoint~\cite{barrierpoint}, TaskPoint~\cite{taskpoint}, and LoopPoint~\cite{looppoint,looppoint-extension} select regions using program structure, while Chronos~\cite{chronos} uses time-based full-system sampling. These approaches require representative coverage of the original execution; \simname{} accelerates replay across the recorded execution and can complement such work reduction.

\subsection{Core Timing Models}
\label{sec:priorcore}

Core timing models differ in how they advance execution and enforce pipeline constraints. \emph{Cycle-driven} models such as gem5's out-of-order CPU~\cite{gem5} explicitly advance pipeline state each cycle, representing detailed interactions at substantial evaluation cost. Sniper's \emph{interval simulation}~\cite{interval,sniper} reduces this cost by analytically estimating instruction overlap and miss penalties, leaving some operation--resource interactions implicit~\cite{iw-centric}.

\emph{Instruction-driven} models such as ZSim~\cite{zsim} and Prophet~\cite{prophet} evaluate instruction and resource constraints without repeated cycle sweeps. Calipers~\cite{calipers} provides a related constraint-based formulation through instruction-event graphs and critical-path analysis. Instruction-level evaluation retains dependencies and resource constraints at relatively low cost, but its fidelity still depends on the scheduling and completion relationships represented. \simname{} builds on this balance, explicitly connecting readiness-driven issue, resource occupancy, and completion feedback to address the discrepancies examined in \S\ref{sec:motivation-core}.

\emph{Machine-learning} models offer another way to reduce timing evaluation cost. SimNet~\cite{simnet} and TAO~\cite{tao} predict instruction timing, while Concorde~\cite{concorde} predicts region CPI from analytical component estimates. Their fidelity depends on training coverage and generalization. \simname{} retains explicit timing rules for instruction scheduling and core--memory interactions rather than learning their combined effects.

\subsection{Parallel Simulation}
\label{sec:priorparallel}

\emph{Spatial decomposition} maps target cores or components to host workers, as in Graphite~\cite{graphite}, ZSim~\cite{zsim}, parti-gem5~\cite{partigem5}, and SplitSim~\cite{splitsim}. Coordinated progress maintains inter-component timing interactions but incurs communication and synchronization costs. Staged coordination in LEGOSim~\cite{legosim} and causal-frontier synchronization in FCS~\cite{fcs} reduce these costs; available concurrency remains tied to the target hardware structure, limiting host utilization for small target systems.

\emph{Temporal decomposition} exposes work at different positions within one execution. Early parallel trace processing~\cite{traceparallel1997}, DiST~\cite{girbal2003dist}, and GPU sub-trace simulation by Pandey et al.~\cite{pandey2022gpu} exploit this opportunity within single-core streams. For parallel programs, Barrier-Interval Time-Parallelism~\cite{barrier-interval} and ChunkedTejas~\cite{chunkedtejas} construct system-wide intervals using global barriers and reference progress or stable-phase predictions, respectively. These methods require corresponding partition boundaries across threads, constraining partition flexibility. Within each interval, detailed simulation remains coupled across cores to resolve their interactions.
Unlike ChunkedTejas's joint multicore progression within intervals, \simname{} combines independent core-local replay with dependency-based time reconstruction.

Exposing parallelism through independent regional replay requires both dependency capture and timing composition. MUSA~\cite{musa} evaluates MPI-rank computation phases in parallel and combines their durations through communication replay, retaining multicore simulation within each phase. For synchronization capture, SynchroTrace~\cite{synchrotrace} and ElasticSimMATE~\cite{elasticsimmate} instrument synchronization APIs or runtimes, while HETSIM~\cite{hetsim} uses explicitly specified primitives. These mechanisms tie explicit synchronization recognition to supported software interfaces. \simname{} uses dependencies recovered from instruction-level release--acquire handoffs, reducing reliance on library-specific tracing, and aggregates independently computed segment durations through a DAG. This supports fine-grained parallel replay of core-local regions without globally aligned boundaries or joint multicore advancement, with approximations for entry states and cross-core interference.

Finer regions make entry-state reconstruction more frequent. Perelman et al.~\cite{perelman2003simpoint} retain stale state, while Haskins and Skadron~\cite{haskins2005warmup} combine retention with recent-history warmup, preserving potentially useful state beyond a short warmup window. MTR~\cite{mtr} and TraceRTL~\cite{tracertl} reduce reconstruction work through compressed memory history and warmup deduplication, respectively. \simname{} combines retained state and warmup with synchronization-guided cache eviction to approximate local history and cross-core cache effects without predecessor replay results.

\section{Motivation}
\label{sec:motivation}

\subsection{Accurate Instruction-Driven Timing}
\label{sec:motivation-core}
\label{sec:priorcore-fidelity}

Instruction-driven models avoid repeated pipeline updates, but simplified scheduling can change how operations compete for resources. 
When a younger operation is ready while an older one waits for operands, trace-order updates may expose it to the older operation's effects before they occur in target time.

\begin{figure}[t]
  \centering
  \includegraphics[width=0.85\columnwidth]{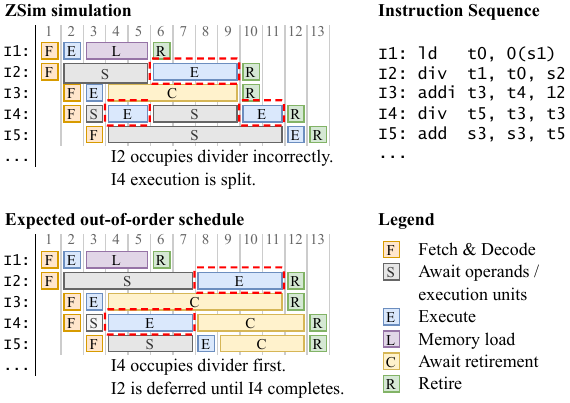}
  \caption{Examples of timing distortions from simplified scheduling in ZSim's instruction-driven model.}
  \label{fig:instruction-violation}
\end{figure}

Figure~\ref{fig:instruction-violation} illustrates this problem. At cycle~4, \texttt{I2} is waiting for the result of \texttt{I1}, while \texttt{I4} is ready to issue. Because ZSim~\cite{zsim} processes instructions in program order, it reserves the divider for \texttt{I2} during cycles~6--9 before scheduling \texttt{I4}. This reservation splits the modeled execution of \texttt{I4} into two intervals, distorting the writeback and retirement timing and affecting other instructions. In the target out-of-order execution, \texttt{I4} would occupy the divider first, and \texttt{I2} would wait until \texttt{I4} completes.

Scheduling and memory service interact because completions determine operand readiness and release finite resources, while long-latency responses can delay retirement and instruction admission.
Scheduling, in turn, determines request arrival times. Translation delays and page-table traffic participate in this feedback, so their effects extend beyond the latency of individual accesses.

\begin{figure}[t]
    \centering
    \includegraphics[width=\columnwidth]{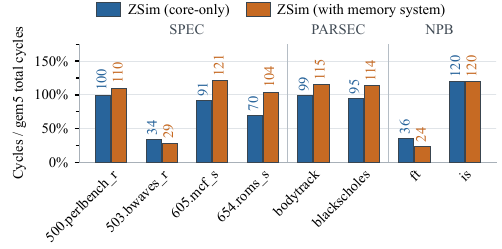}
    \caption{ZSim cycles under fixed-latency memory and native memory systems, normalized to the corresponding gem5
    reference in each setting.}
    \label{fig:core-ablation}
\end{figure}

To assess the prediction impact of these timing simplifications, we extract ZSim's core and memory timing models~\cite{zsim} and replay instruction traces recorded from gem5~\cite{gem5} to hold the execution path fixed within each comparison. These controlled experiments use four-wide x86 configurations and measure $10^6$ instructions after $10^5$ instructions of warmup. Figure~\ref{fig:core-ablation} reports fixed-latency and native-memory experiments, each normalized to its corresponding gem5 reference. Fixed-latency memory suppresses detailed memory-hierarchy effects to expose differences in core timing: ZSim predicts only 34.2\% of gem5's cycles for \texttt{503.bwaves\_r} and 35.7\% for \texttt{FT}. Such deviations motivate more faithful enforcement of instruction readiness and resource constraints within an efficient instruction-driven model.

The native-memory experiments additionally expose the influence of memory modeling and its interaction with core execution. For \texttt{605.mcf\_s} and \texttt{blackscholes}, ZSim predicts 121.2\% and 114.1\% of the respective native-memory gem5 references. Their fixed-latency results are 91.3\% and 94.8\% of the separate fixed-latency references. The differing error patterns reinforce the need to model memory service and its feedback into instruction scheduling, beyond improving the core model in isolation. These observations motivate \simname{} to combine explicit readiness-driven out-of-order issue with lightweight address-translation and memory-hierarchy models, coupled through request and completion timing to preserve these interactions while retaining instruction-driven efficiency.

\subsection{Synchronization-Guided Trace Decomposition}
\label{sec:motivation-parallel}

Temporal decomposition exposes additional simulation concurrency, but existing coarse-grained approaches~\cite{barrier-interval, chunkedtejas} retain joint replay across cores within each interval and thus require specific program structures or advance knowledge of relative thread progress. Independent replay of finer-grained per-core regions could expose more parallel work. However, simply partitioning per-core traces (e.g., by instruction count) leaves two problems unresolved. First, these cuts may miss cross-core dependency endpoints and fail to preserve their precedence. Second, segment overlap across cores is undetermined under such partitioning, preventing faithful modeling of cross-core interference, especially cache effects that depend on precise access interleaving.

\begin{figure}[t]
  \centering
  \includegraphics[width=0.85\columnwidth]{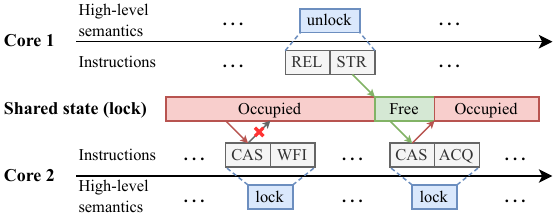}
  \caption{High-level lock synchronization implemented through an instruction-level release--acquire handoff.}
  \label{fig:synchronization}
\end{figure}

Synchronization provides a structure for recovering these relationships. Parallel programs use mechanisms such as locks and barriers to establish the ordering and visibility guarantees required for correct execution. At a lower level, these mechanisms are constructed of shared-data communication through release--acquire handoffs: a producer publishes updates through a release, and a consumer completes a corresponding acquire before using them~\cite{posix2018,hennessy2017computer}. These handoffs are implemented through memory accesses, atomic operations, and ordering primitives~\cite{riscv-a-extension,arm-armv8,intel-sdm}. Figure~\ref{fig:synchronization} illustrates how a high-level \texttt{unlock()} maps to a release-ordered store, while successful completion of \texttt{lock()} corresponds to an atomic acquisition. This instruction-level structure provides a common observation point across synchronization implementations.

Matched handoffs provide a basis for addressing both problems. They identify segment boundaries and cross-core precedence edges that, together with same-core order, form a dependency DAG for combining independently computed segment durations. Meanwhile, such handoffs associate preceding producer writes with subsequent consumer execution, allowing their cache-invalidating effects to be approximated at the boundary.
Synchronization therefore supports reconstruction of cross-core temporal relationships and estimation of handoff-related cache interference while local execution times are computed independently.

\section{Design of \simname{}}
\label{sec:method}

\subsection{Overview}
\label{sec:overview}

Building on the observations in \S~\ref{sec:motivation}, \simname{} uses synchronization-guided temporal decomposition to partition execution traces for parallel replay and aggregate segment durations through a dependency DAG. An instruction-driven core model with explicit out-of-order issue and coupled translation and memory timing provides fast and accurate estimates of segment-local execution cycles.

\begin{figure}[t]
  \centering
  \includegraphics[width=0.85\columnwidth]{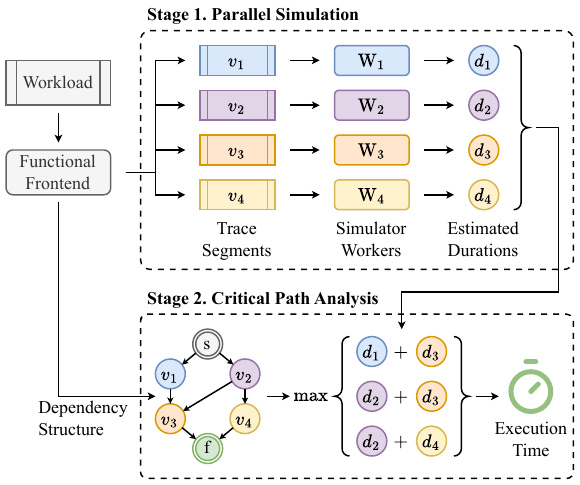}
  \caption{The \simname{} workflow, from execution traces to execution-time prediction.}
  \label{fig:overview}
\end{figure}

Figure~\ref{fig:overview} shows the \simname{} workflow for predicting the execution time of a target workload. A functional frontend, such as a functional simulator or a proxy machine, executes the workload and generates an execution trace containing the information listed in Table~\ref{tab:gipsim-input}. Before independent parallel replay begins, \simname{} partitions the per-core traces into segments and constructs a DAG $G=(V,E)$ whose edges encode program order within each core and synchronization dependencies across cores.

\begin{table}[t]
    \centering
    \caption{Trace information required by \simname{}.}
    \label{tab:gipsim-input}

    \small
    \setlength{\tabcolsep}{4pt}
    \renewcommand{\tabularxcolumn}[1]{m{#1}}
    \begin{tabularx}{\columnwidth}{lY}
        \toprule
        Record type             & Required information                                                      \\
        \midrule
        Instructions            & Instruction addresses, \uop{} decomposition, and branch outcomes          \\
        Memory accesses         & Access addresses, sizes, and read/write directions                        \\
        Synchronization         & Matched synchronization endpoints                                         \\
        Address translation     & Virtual-to-physical mappings, page sizes, and translation invalidations   \\
        \bottomrule
    \end{tabularx}
\end{table}

In Stage~1, host workers compute segment durations independently in parallel (\S~\ref{sec:parallel}), with our instruction-driven core model and its coupled memory system (\S~\ref{sec:core}). For segment $v$, we compute
\begin{equation}
  d_v=\mathcal{M}(\mathcal{T}_v,\theta;\widehat{x}_v,\widehat{h}_v),
  \label{eq:local-duration}
\end{equation}
where $\mathcal{T}_v$ is the segment's trace, $\theta$ specifies the target configuration, $\widehat{x}_v$ approximates its entry state, and $\widehat{h}_v$ estimates shared-resource interference. History reconstruction and interference estimation provide the context needed for independent replay without requiring predecessor timing results (\S~\ref{sec:warmup}). Each duration $d_v$ is expressed in target cycles.

In Stage~2, segment durations become node weights in $G$. Critical-path analysis aggregates these weights under the ordering constraints to predict program completion time (\S~\ref{sec:critical-path}).

\subsection{Trace Decomposition and Task Construction}
\label{sec:parallel}
\label{sec:extraction}
\label{sec:segmentreplay}

\simname{} partitions each core's trace at synchronization endpoints and constructs a directed acyclic graph $G=(V,E)$ with the resulting segments as vertices. The graph contains two edge types:
\begin{equation}
  E=E_{\mathrm{seq}}\cup E_{\mathrm{sync}}.
  \label{eq:edges}
\end{equation}
Edges in $E_{\mathrm{seq}}$ connect consecutive segments on the same core. Edges in $E_{\mathrm{sync}}$ connect a producer's release endpoint to the consumer segment constrained by the corresponding successful acquisition, as illustrated in Figure~\ref{fig:synchronization}; unsuccessful attempts do not establish these handoff edges.

To reduce heavy context reconstruction overheads for independent replay when synchronization is frequent, \simname{} aggregates consecutive segments along each core's trace into a \emph{task} with $10^6$ instructions. Segments crossing task boundaries are split, introducing additional edges in $E_{\mathrm{seq}}$.

Workers claim tasks dynamically and replay all constituent segments consecutively, recording each segment's duration separately. Microarchitectural state is inherited across segments within a task, with only synchronization-related cache line evictions applied at the relevant boundaries.

\subsection{Context for Independent Replay}
\label{sec:shared}

\simname{} uses multiple simulator workers to replay tasks independently, with each task maintaining its own timing context. To simulate segments in a well-approximated context, \simname{} reconstructs entry states and models interference from other segments.

\noindent
\textbf{Local history.}
\label{sec:warmup}
Workers reuse \emph{stale microarchitectural state} retained from earlier replay to avoid repeated cold initialization. At task entry, a warmup of approximately $10^4$ preceding instructions from the same target core adapts caches, TLBs, predictors, and prefetchers to reflect recent execution. Warmup cycles are excluded from measured durations. Subsequent segments within the task inherit the context without additional warmup, avoiding repeated state reconstruction at every synchronization boundary.

\noindent
\textbf{Synchronization-related cache state.}
\label{sec:sync-invalidation}
To approximate handoff-related cache effects, \simname{} collects the physical cache lines written within approximately $10^4$ instructions preceding each matched release. Before the consumer enters the corresponding segment, these lines are evicted from its local data-cache hierarchy, with tags to indicate their remote origin. Subsequent accesses matching these tags follow a timing path that approximates retrieval from another core's cache rather than from DRAM. For successive handoffs from the same producer core, a high-water mark tracks the processed position in the write history so that invalidations are applied only to newly covered writes.

\noindent
\textbf{Incremental DRAM queueing.}
\label{sec:dram-correction}
\simname{} approximates DRAM contention using a simplified queueing model. During preprocessing, it traverses the dependency DAG using segment instruction counts as proxy durations to estimate $\alpha_v$, the average number of active target cores during segment $v$. During replay, local utilization $\rho_{\mathrm{local}}$ is estimated from recent DRAM request density. Scaling this utilization by $\alpha_v$ approximates aggregate pressure from the active cores. Using the M/D/1~\cite{data-networks} mean queueing expression,
\begin{equation}
  W(\rho)=\frac{\rho\,\tau_s}{2(1-\rho)},\qquad 0\leq\rho<1,
  \label{eq:queue-correction}
\end{equation}
where $\tau_s$ is an effective request service time. For $\alpha_v\rho_{\mathrm{local}}<1$, the baseline incremental wait is
\begin{equation}
  \Delta W_v=\max\!\left\{0,\,
    W(\alpha_v\rho_{\mathrm{local}})-W(\rho_{\mathrm{local}})\right\}.
  \label{eq:extra-wait}
\end{equation}
A finite cap bounds the waiting-time estimates near and above saturation.
These equations describe an analytical approximation, not an assumption that DRAM traffic exactly follows an M/D/1 process. Scaling local intensity cannot reproduce heterogeneous per-core traffic or globally ordered bank conflicts, but estimates contention without per-request coordination.

\subsection{Instruction-Driven Core and Memory Timing}
\label{sec:core}

\begin{figure}[t]
  \centering
  \includegraphics[width=\columnwidth]{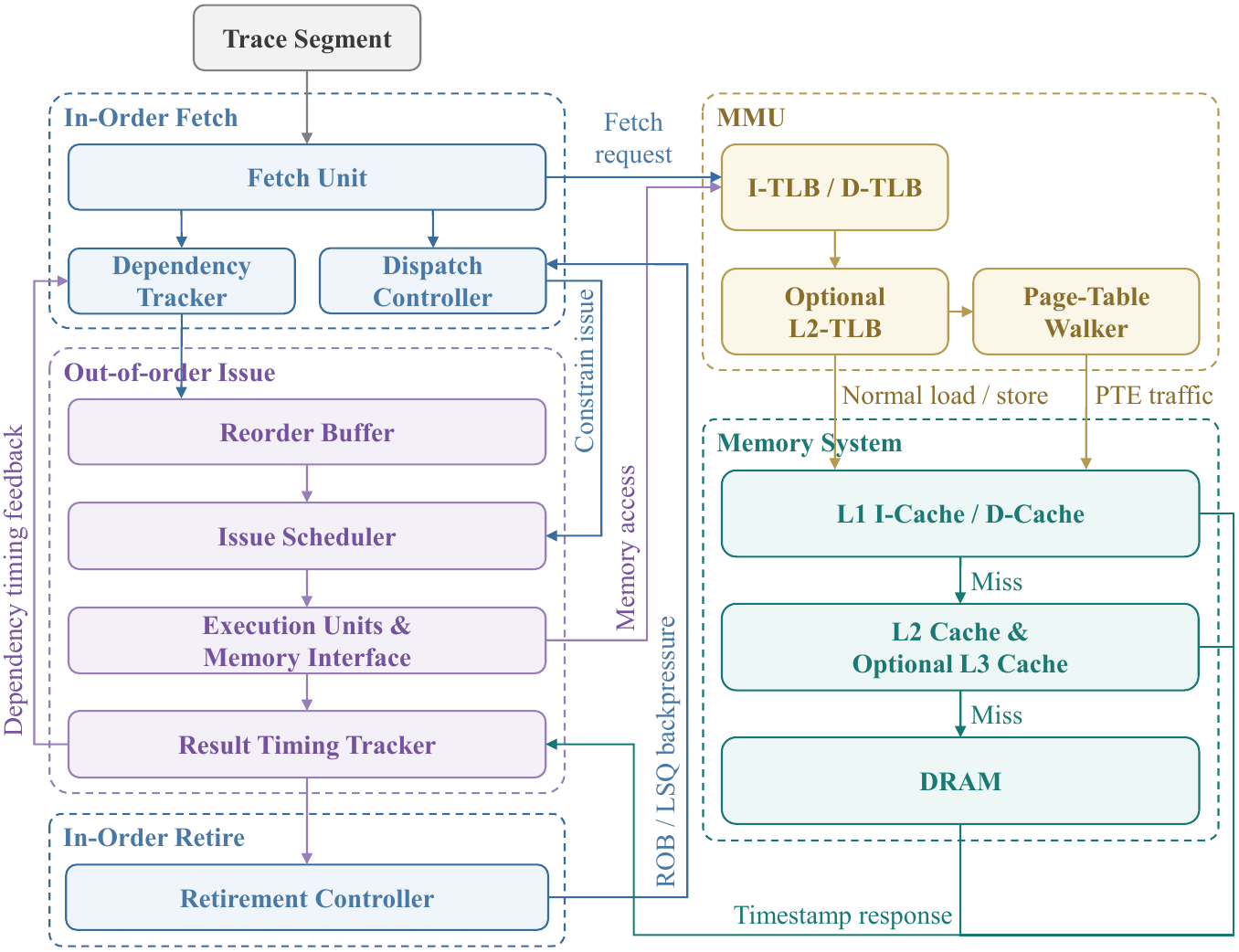}
  \caption{The simplified single-core system model in \simname{}.}
  \label{fig:single-core}
\end{figure}

\simname{} models an out-of-order pipeline with \emph{in-order fetch and dispatch}, \emph{out-of-order issue and execution}, and \emph{in-order commit} (retirement), as shown in Figure~\ref{fig:single-core}. Dependencies, resource occupancy, and completion timestamps couple these stages. A timestamp-based memory model for address translation, cache access and DRAM service timing is coupled to the core model with request completions feeding back into instruction readiness, resource availability, and pipeline progress.

\noindent
\textbf{In-order fetch and dispatch.}
The pipeline frontend consumes the instruction trace in program order, accounting for fetch width, instruction-cache access, and rename latency. Dispatch is constrained by finite reorder-buffer (ROB), load-queue (LQ), and store-queue (SQ) capacities and stalls when required entries are unavailable. Register producer--consumer relationships are established for the admitted operations. This stage also models branch prediction, including branch direction and target prediction. The predictor includes a branch target buffer (BTB), return-address stack (RAS), and indirect branch predictor. Mispredictions occur when the prediction differs from the corresponding outcome in the trace, stalling subsequent fetching for a recovery delay.

\noindent
\textbf{Out-of-order issue and execution.}
Dispatched \uop{}s compete for execution according to operand readiness and resource availability. For operation $i$, let $a_i$ be its dispatch time, $\mathcal{D}_i$ its data dependencies, and $b_{j\rightarrow i}$ the time when producer $j$'s result becomes available to it. Its earliest data-ready time is
\begin{equation}
  r_i=\max\left(a_i,\max_{j\in\mathcal{D}_i}b_{j\rightarrow i}\right),
  \label{eq:ready}
\end{equation}
with $r_i=a_i$ for an empty dependency set. Dependency tracking covers register dependencies, intra-instruction \uop{} dependencies, and forwarding delays.

A dependency counter tracks producers whose result-availability times have not yet been computed. Once all required availability times are known, the operation is inserted into a priority queue implemented with timestamp buckets. Ready operations are issued in nondecreasing order of their ready cycles, subject to the per-cycle issue-width limit. The completion cycles of the execution, load, store, and writeback stages are then computed directly using the start cycles plus the latencies of the corresponding stage. An operation with an unresolved ready cycle must wait for its producers to complete execution, so it cannot become ready earlier than the earliest known ready cycle awaiting scheduling.

\noindent
\textbf{In-order commit and backpressure.} Commit proceeds only from the ROB head and is constrained by retirement bandwidth. Younger operations that finish early remain in the ROB until older operations retire. \simname{} models this behavior by retaining completed operations in its ROB slot array and retiring them in order, subject to the commit width. A long-latency operation at the head can therefore block retirement, fill the ROB, and block the pipeline frontend. This preserves feedback from out-of-order execution through in-order commit to in-order dispatch.

\noindent
\textbf{Translation and memory timing.}
\simname{} models a memory management unit (MMU) for address-translation timing and a timestamp-based memory system for request service. The MMU includes separate instruction and data TLBs, an optional shared second-level TLB, and page-table walkers with finite concurrency. The memory system comprises multilevel caches with miss-handling resources, prefetching, and writeback support, together with a DRAM model that tracks row state, bank and channel contention, and write buffering.

For instruction fetches and data accesses, \simname{} first obtains the virtual and physical addresses from the trace and determines translation timing through the MMU. A first-level TLB hit is modeled as overlapped with cache access and adds no separate delay, while a miss initiates a second-level lookup when configured. An unresolved translation triggers a page-table walk, whose dependent reads traverse the configured memory path. Once translation timing has been determined, the original access is timed through the cache hierarchy and, when required, DRAM. The resulting completion cycle is returned to the core model, determining instruction-fetch timing and data-request completion. For memory stores, cache writes occur after \uop{} retirement, so their timing effects primarily stem from continued SQ occupancy and the resulting backpressure on store dispatch.

\subsection{Time Reconstruction}
\label{sec:critical-path}

Local replay determines each segment's duration $d_v$ under the pipeline and memory constraints above. After replay, these durations become node weights in the dependency graph, which determines their placement on the target timeline. Let $\mathrm{pred}(v)$ denote the immediate predecessors of $v$. A topological traversal computes its start and finish times:
\begin{equation}
  \begin{aligned}
    S_v&=\max\left(\{0\}\cup
                   \{F_u\mid u\in\mathrm{pred}(v)\}\right),\\
    F_v&=S_v+d_v.
  \end{aligned}
  \label{eq:start}
\end{equation}
The predicted completion time is
\begin{equation}
  \widehat{T}=\max_{v\in V}F_v.
  \label{eq:finish}
\end{equation}
If a consumer would otherwise start before its synchronization predecessor finishes, the corresponding edge postpones its start; otherwise, that edge adds no waiting. The traversal takes $O(|V|+|E|)$ time.

\begin{figure*}[t]
  \centering
  \includegraphics[width=\linewidth]{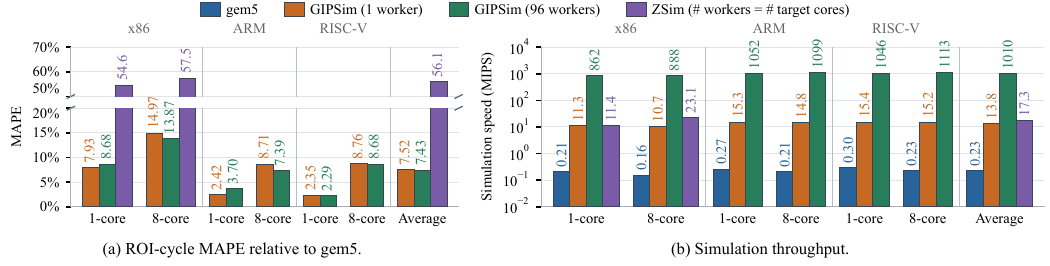}
  \caption{Timing accuracy and simulation throughput across ISAs and target core counts.}
  \label{fig:eval-summary}
\end{figure*}

\section{Evaluation}
\label{sec:evaluation}

We evaluate \simname{} in terms of timing accuracy and simulation efficiency. We first summarize overall results, then examine per-workload timing accuracy, simulation speed and scalability, and accuracy across hardware configurations. We subsequently report trace acquisition costs and evaluate the sensitivity to warmup and cache-eviction history lengths.

\subsection{Experimental Setup}
\label{sec:eval-setup}

\noindent
\textbf{Host platform.}
The host platform is a dual-socket Intel Xeon Platinum 8558P server with 96 physical cores, two NUMA nodes, and 256\,GiB memory. One hardware thread is enabled per core, and the frequency policy is configured at 3.1\,GHz. The server runs Ubuntu 24.04.3 with Linux 6.8.0.

\noindent
\textbf{Target systems.}
We evaluate \simname{} on target systems with x86, ARM, and RISC-V ISAs, each configured with one and eight cores. Table~\ref{tab:eval-target} summarizes the target configurations.

\begin{table}[t]
    \centering
    \caption{Target system configuration.}
    \label{tab:eval-target}

    \small
    \setlength{\tabcolsep}{4pt}
    \renewcommand{\tabularxcolumn}[1]{m{#1}}
    \begin{tabularx}{\columnwidth}{lY}
        \toprule
        Component & Configuration \\
        \midrule
        ISA / cores             & x86-64, AArch64, RV64GC; 1 or 8 cores \\
        Pipeline                & 8-wide out-of-order core \\
        Resource constraints    & ROB: 192; IQ: 64; LQ/SQ: 32/32 \\
        Branch predictor        & Tournament; 4096-entry BTB; 16-entry RAS \\
        L1I / L1D               & 32\,KiB each, 8-way; 64-byte lines; tag/data latencies: 1/1 cycles \\
        L2                      & 256\,KiB, 16-way, private; tag/data latencies: 10/10 cycles \\
        Prefetching             & Degree-4 stride prefetching for data accesses \\
        Main memory             & 3\,GiB, single-channel DDR4-3200 \\
        DRAM organization       & Two ranks; 16 banks/rank; FR-FCFS reference \\
        \bottomrule
    \end{tabularx}
\end{table}

\begin{table}[t]
    \centering
    \caption{Target workloads.}
    \label{tab:target-workload}

    \small
    \setlength{\tabcolsep}{4pt}
    \begin{tabular*}{\columnwidth}{@{\extracolsep{\fill}\hspace{\tabcolsep}}lll}
        \toprule
        Benchmark Suite & Workload & Abbr. \\
        \midrule
        \multirow{4}{*}{SPEC CPU 2017}
            & \texttt{500.perlbench\_r:test}    & \texttt{PB} \\
            & \texttt{503.bwaves\_r:test}       & \texttt{BW} \\
            & \texttt{605.mcf\_s:test}          & \texttt{MC} \\
            & \texttt{654.roms\_s:test}         & \texttt{RO} \\
        \midrule
        \multirow{2}{*}{PARSEC}
            & \texttt{bodytrack:test}           & \texttt{BT} \\
            & \texttt{blackscholes:simlarge}    & \texttt{BS} \\
        \midrule
        \multirow{2}{*}{NPB}
            & \texttt{FT:S}                     & \texttt{FT} \\
            & \texttt{IS:A}                     & \texttt{IS} \\
        \bottomrule
    \end{tabular*}
\end{table}

\noindent
\textbf{Baseline simulators.}
We use gem5~\cite{gem5} version 25.0.0.1 as the timing reference, with O3CPU simulating the region of interest (ROI) in full-system mode. ZSim~\cite{zsim} provides an additional baseline for simulation throughput. Because ZSim only models user-space execution, its cycle counts exclude kernel execution and may therefore be lower than full-system estimates. Prophet~\cite{prophet} is closely related to \simname{}, but our reproduction attempts yielded simulation throughput below 0.1~MIPS and execution-time prediction errors as high as 80\%. We exclude it from application-level comparisons. We discuss ChunkedTejas~\cite{chunkedtejas} qualitatively in \S\ref{sec:priorparallel}, as we could not locate a public implementation of its chunking extensions.

\noindent
\textbf{Workloads and ROI.}
All configurations use eight common workloads listed in Table~\ref{tab:target-workload}. We use test inputs for SPEC CPU 2017~\cite{spec2017} and PARSEC~\cite{parsec}, and class S for NPB~\cite{npb}, with two exceptions: \texttt{blackscholes} uses \texttt{simlarge}, and \texttt{IS} uses class A, to obtain sufficiently long execution traces.
Each guest workload is configured with a degree of parallelism matching the target core count. The ROI begins when the workload starts and ends when it terminates or the aggregate instruction count across all target cores reaches $10^9$ for single-core targets or $8\times10^9$ for eight-core targets, whichever occurs first. Only retired instructions are counted; instructions executed speculatively and subsequently squashed are excluded.

\begin{figure*}[t]
  \centering
  \includegraphics[width=\linewidth]{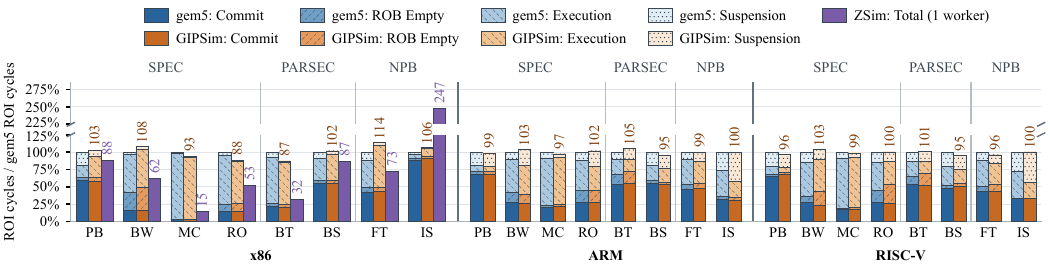}
  {\small (a) Single-core targets.}
  \par\medskip
  \includegraphics[width=\linewidth]{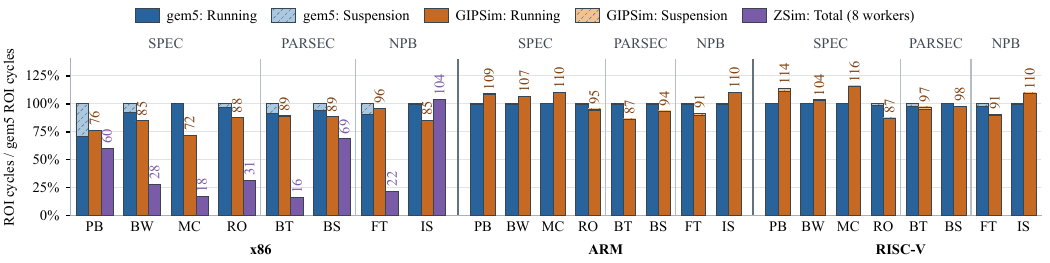}
  {\small (b) Eight-core targets.}
  \caption{Per-workload simulated cycles normalized to the corresponding gem5 ROI cycles. \simname{} uses one host worker, while ZSim uses one host worker per target core.}
  \label{fig:eval-default-accuracy}
\end{figure*}

\begin{figure*}[t]
  \centering
  \includegraphics[width=\linewidth]{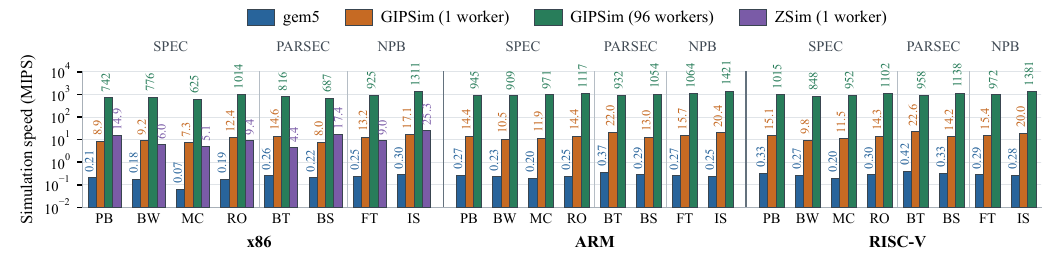}
  {\small (a) Single-core targets.}
  \par\medskip
  \includegraphics[width=\linewidth]{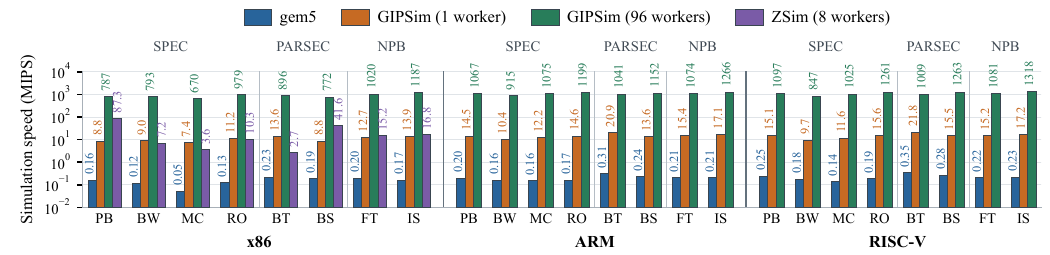}
  {\small (b) Eight-core targets.}
  \caption{Per-workload simulation throughput with one and 96 host workers, together with the gem5 reference measurements.}
  \label{fig:eval-throughput}
\end{figure*}

\begin{figure}[t]
  \centering
  \includegraphics[width=\linewidth]{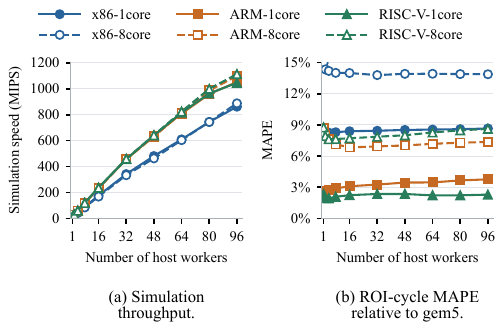}
  \caption{Simulation throughput and timing error of \simname{} as the number of host workers increases.}
  \label{fig:eval-scaling}
\end{figure}

\noindent
\textbf{Evaluation Metrics.}
We evaluate simulation accuracy using cycle-count error relative to gem5 and report the MAPE across workloads, with each workload receiving equal weight. Simulation throughput is measured in MIPS, based on the number of ROI instructions and total host-side simulation time. The total host-side simulation time includes startup, replay, and post-processing.

\subsection{Overall Results}
\label{sec:eval-results}
\label{sec:eval-overall}

Figure~\ref{fig:eval-summary} summarizes timing accuracy and simulation throughput across the six ISA and core-count combinations described in \S~\ref{sec:eval-setup}. ROI-cycle MAPE averages 7.52\% with one host worker and 7.43\% with 96 workers. Averaged arithmetically across workloads and configurations, simulation throughput increases from 13.8 to 1{,}010 MIPS, compared with 0.23 MIPS for gem5 and 17.3 MIPS for ZSim. These throughput measurements include initialization, replay, and post-processing time but exclude trace collection. The ZSim mean covers only the x86 targets.

\subsection{Timing Accuracy}
\label{sec:eval-accuracy}

Figure~\ref{fig:eval-default-accuracy} presents per-workload accuracy for \simname{} using one host worker. For single-core targets, ROI-cycle MAPE is 7.93\%, 2.42\%, and 2.35\% for x86, ARM, and RISC-V, respectively. The cycle breakdowns additionally expose differences in the Commit, ROB Empty, and Execution components, allowing local timing behavior to be examined alongside total-cycle error. The larger x86 deviations include $+13.77\%$ for FT and $-12.68\%$ for BT.

For eight-core targets, ROI-cycle MAPE is 14.97\%, 8.71\%, and 8.76\% for x86, ARM, and RISC-V, respectively. The breakdown separates running cycles from intervals during which all target cores are suspended. For example, x86 PB has a total-cycle error of $-23.51\%$, compared with $+7.63\%$ after excluding all-core suspension. Suspension accounts for 28.95\% of its gem5 ROI cycles.
This case illustrates how suspension-duration differences can dominate total-cycle error despite a substantially closer running-cycle prediction.

\subsection{Speed and Scalability}
\label{sec:eval-speed}

\begin{figure*}[t]
  \centering
  \includegraphics[width=\linewidth]{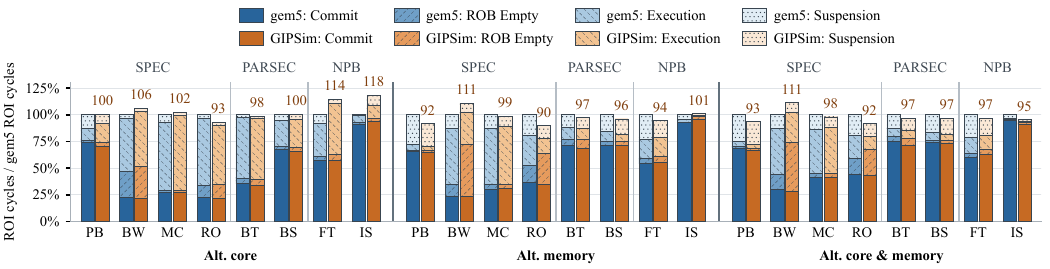}
  {\small (a) Single-core targets.}
  \par\medskip
  \includegraphics[width=\linewidth]{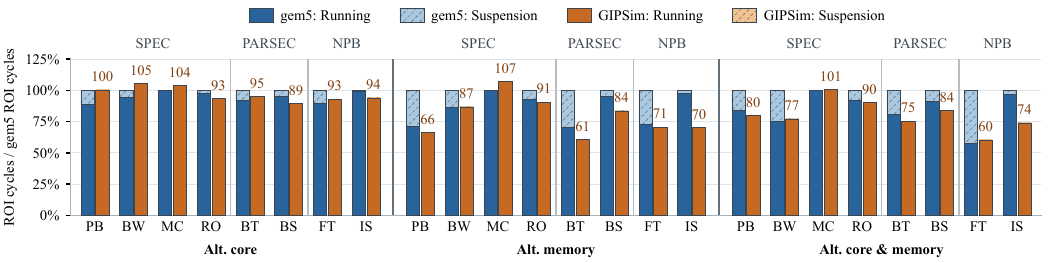}
  {\small (b) Eight-core targets.}
  \caption{Timing accuracy across x86 hardware configurations with one host worker.}
  \label{fig:eval-cross-config}
\end{figure*}

Figure~\ref{fig:eval-throughput} shows simulation throughput across workloads. With 96 host workers, average simulation throughput across workloads reaches 862, 1{,}052, and 1{,}046 MIPS for single-core x86, ARM, and RISC-V targets, respectively; the corresponding eight-core results are 888, 1{,}099, and 1{,}113 MIPS. The single-core results demonstrate that temporal decomposition exposes substantial host parallelism along an individual target core's execution timeline.

Figure~\ref{fig:eval-scaling}(a) varies host worker count independently of target core count. At 16 workers, throughput improves by approximately 15.4--15.8$\times$ over one worker.
At 96 workers, speedup is 68.1--83.0$\times$ across the six systems, with 68.1--76.0$\times$ for single-core targets.

Initialization and post-processing become more visible as parallel replay accelerates. At 96 workers, these stages account for 3.3--4.8\% of total simulation time. The replay stage alone achieves 70.6--86.0$\times$ speedup, compared with 68.1--83.0$\times$ for the complete backend. These measurements distinguish the scalability of detailed replay from the serial work surrounding it.

Figure~\ref{fig:eval-scaling}(b) places these throughput gains alongside their accuracy implications. Averaged across the six systems, ROI-cycle MAPE remains between 7.01\% and 7.52\% throughout the worker-count sweep. Individual configurations exhibit different error trends as replay contexts change, but the aggregate results show limited accuracy variation over the tested range of host parallelism.

\subsection{Accuracy Across Hardware Configurations}
\label{sec:eval-cross-config}

We evaluate three additional x86 configurations that vary core and memory-system parameters separately or in combination, denoted \emph{Alt.\ core}, \emph{Alt.\ memory}, and \emph{Alt.\ core \& memory}. Each is evaluated with one and eight target cores. Relative to Table~\ref{tab:eval-target}, \emph{Alt.\ core} uses four-wide cores with 224-entry ROBs. \emph{Alt.\ memory} adds a 2\,MiB shared L3 and instruction/data page-walk caches of 1\,KiB each, and sets the L1/L2/L3 tag and data latencies to 3/14/67 cycles. \emph{Alt.\ core \& memory} combines these changes.

Figure~\ref{fig:eval-cross-config} evaluates prediction accuracy under these parameter changes. For single-core targets, ROI-cycle MAPE is 6.37\%, 5.49\%, and 5.31\% for \emph{Alt.\ core}, \emph{Alt.\ memory}, and \emph{Alt.\ core \& memory}, respectively. The corresponding eight-core errors are 5.60\%, 22.24\%, and 19.98\%, showing greater sensitivity in the parallel executions of \emph{Alt.\ memory} and \emph{Alt.\ core \& memory}.

The cycle breakdowns reveal a substantial contribution from suspension-duration differences in these cases. Excluding all-core suspension reduces eight-core MAPE to 9.33\% for \emph{Alt.\ memory} and 6.73\% for \emph{Alt.\ core \& memory}. For example, FT under the eight-core \emph{Alt.\ core \& memory} configuration has a total-cycle error of $-39.59\%$, compared with $+4.05\%$ for running cycles; all-core suspension occupies 42.29\% of the gem5 ROI. These results show the importance of distinguishing active-execution prediction from suspension-duration modeling when evaluating accuracy across hardware configurations.

\subsection{Trace Generation and Storage Overhead}
\label{sec:eval-trace-cost}

\begin{table}[t]
    \centering
    \caption{Trace acquisition and storage overheads, with means in square brackets.}
    \label{tab:eval-trace-cost}

    \small
    \setlength{\tabcolsep}{4pt}
    \begin{tabular*}{\columnwidth}{@{\extracolsep{\fill}\hspace{\tabcolsep}}lrrr}
        \toprule
        Target & Instructions (\(\times10^9\)) & Capture (s) & Trace (GiB) \\
        \midrule
        x86 / 1-core  & 0.40--1.00 [0.89] & 80--191 [152]  & 8.5--29.5 [22.4]   \\
        ARM / 1-core  & 0.34--1.00 [0.87] & 83--166 [138]  & 7.6--29.0 [20.9]   \\
        RV64 / 1-core & 0.35--1.00 [0.87] & 35--112 [87]   & 7.7--27.5 [20.9]   \\
        x86 / 8-core  & 0.40--8.00 [5.08] & 59--1296 [306] & 8.7--236.1 [136.3] \\
        ARM / 8-core  & 0.34--8.00 [4.95] & 64--896 [245]  & 7.7--232.3 [124.2] \\
        RV64 / 8-core & 0.36--8.00 [5.03] & 21--871 [198]  & 7.8--220.5 [122.9] \\
        \bottomrule
    \end{tabular*}
\end{table}

Table~\ref{tab:eval-trace-cost} quantifies the trace acquisition and storage costs excluded from the backend throughput measurements. For each target configuration, we first average repeated captures of each workload, then report the range and arithmetic mean across the eight workloads. Acquisition time covers the complete QEMU capture process, including guest startup, while storage is measured as the total size of the recorded trace files. Actual ROI instruction counts are included to indicate the amount of execution captured.

Across ISAs, mean acquisition time ranges from 87 to 152 seconds for single-core targets and from 198 to 306 seconds for eight-core targets. The corresponding mean trace sizes are 20.9--22.4\,GiB and 122.9--136.3\,GiB. These costs accompany different amounts of recorded execution: configuration-level mean ROI lengths are 0.87--0.89 billion instructions for single-core targets and 4.95--5.08 billion instructions for eight-core targets.

\subsection{Sensitivity of Replay Context Reconstruction}
\label{sec:eval-context-sensitivity}

We examine how the warmup length $W$ and cache-eviction history length $E$ affect reconstructed replay contexts. The application sweeps reuse a fixed trace for each workload, with one replay per setting. We report mean absolute cycle deviations across workloads and aggregate simulation time.

\begin{figure}[t]
  \centering
  \subfloat[Warmup length.\label{fig:eval-warmup}]{%
    \includegraphics[width=0.85\linewidth]{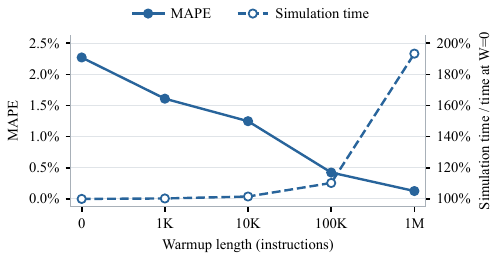}}
  \par\medskip
  \subfloat[Cache-eviction history length.\label{fig:eval-invalidation}]{%
    \includegraphics[width=0.85\linewidth]{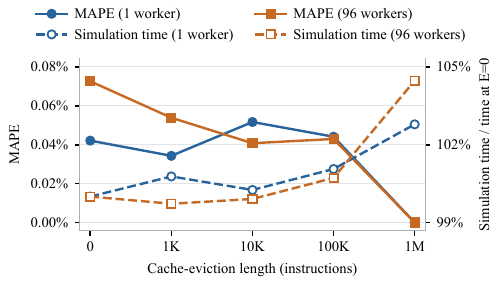}}
  \caption{Sensitivity to the history lengths used for replay context reconstruction.}
  \label{fig:eval-context-sensitivity}
\end{figure}

\noindent\textbf{Warmup length.}
Figure~\ref{fig:eval-warmup} uses the eight single-core x86 workloads with 96 host workers and $W\in\{0,1\mathrm{K},10\mathrm{K},100\mathrm{K},1\mathrm{M}\}$. Cycle deviations are measured against a one-worker replay of the same trace without warmup or periodic task boundaries. This reference preserves microarchitectural history throughout the ROI and approximates uninterrupted simulation, allowing the effect of boundary-state reconstruction to be assessed. Simulation time is normalized to the 96-worker measurement with $W=0$.

Increasing $W$ from zero to 10K reduces mean cycle deviation from 2.27\% to 1.25\%, with a 1.54\% increase in simulation time. A 100K window reduces deviation to 0.42\% at a 10.23\% time increase, while a 1M window reaches 0.13\% at a 93.31\% increase. Longer prefixes improve agreement with continuous replay but repeat more simulation work. We use 10K as the default warmup length to balance this improvement against reconstruction cost.

\noindent\textbf{Cache-eviction history length.}
Figure~\ref{fig:eval-invalidation} uses the eight eight-core x86 workloads with one and 96 host workers and 10K warmup. We vary $E$ over the same five history lengths. Cycle deviations use the corresponding $E=1\mathrm{M}$ replay as the reference, comparing shorter producer histories with the longest window tested.
Simulation time is normalized to $E=0$ for each worker count.

At $E=10\mathrm{K}$, mean cycle deviation is 0.052\% with one worker and 0.041\% with 96 workers. Across all tested settings, deviation remains below 0.08\%. The corresponding simulation-time changes at 10K are $+0.26\%$ and $-0.09\%$ relative to $E=0$; at 1M, time increases by 2.78\% and 4.47\%, respectively. These applications therefore exhibit limited timing sensitivity to the producer-history window. The modest cost growth may partly reflect the high-water mark, which avoids reprocessing overlapping producer-write history.

\begin{figure}[t]
  \centering
  \includegraphics[width=0.85\linewidth]{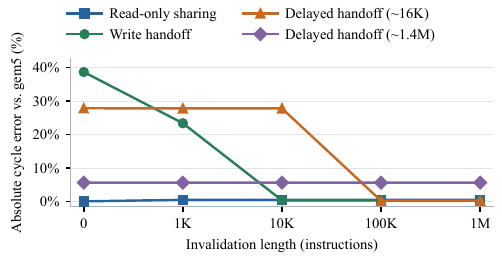}
  \caption{Sensitivity to cache-eviction history length under controlled sharing patterns, relative to gem5.}
  \label{fig:eval-invalidation-microbenchmarks}
\end{figure}

To examine which shared-data updates a bounded window can capture, we additionally use four two-core x86 sharing patterns. Each uses a 16\,KiB shared payload, one host worker, and 10K warmup. Consumers traverse an address-dependent randomized chain to limit memory-level parallelism and sequential prefetching. All window settings for a pattern replay the same gem5 retirement trace, and Figure~\ref{fig:eval-invalidation-microbenchmarks} reports absolute cycle error against the corresponding gem5 ROI.

For immediate write handoff, disabling eviction underestimates cycles by 38.63\%. A 10K window reduces absolute error to 0.37\%, with little further change at longer windows. The read-only pattern remains within 0.09--0.55\% error across the sweep. With a handoff delayed by approximately 16K instructions after the last payload write, the 10K window leaves 27.78\% error, while 100K reduces it to 0.25\%. In the fourth pattern, publication follows the last payload write by approximately 1.4M instructions. Even the largest tested window cannot reach that write, and error remains approximately 5.66\% throughout the sweep.

These results support a 10K default for low-cost reconstruction in the evaluated applications and immediate handoffs. The delayed-sharing cases show that window coverage depends on publication distance, motivating a configurable producer-history length.

\section{Discussion and Limitations}
\label{sec:limitations}

\noindent\textbf{Simulation mode.}
The current implementation of \simname{} is trace-driven. Reusing the same trace across hardware configurations avoids repeated functional execution during timing evaluations, making this approach well suited to design-space exploration. The benefit is limited when evaluating different workloads under a fixed configuration, since each workload requires a separate trace, incurring storage requirements and disk I/O overhead. An execution-driven implementation could reduce these overheads. To implement this extension, the local timing and DAG-based reconstruction mechanisms can be reused, with tasks and synchronization dependencies generated incrementally during functional execution.

\noindent\textbf{Cross-core interactions.}
\simname{}'s independent replay uses synchronization relationships to approximate cross-core data exchange. The current decomposition therefore targets data-race-free programs without false sharing. Data races can make behavior depend on accesses not ordered by synchronization. False sharing can introduce coherence traffic between otherwise independent accesses to different objects on the same cache line. These interactions can affect local timing beyond the recorded synchronization graph and require additional cross-region information.

\noindent\textbf{Suspension time.}
The DAG accounts for synchronization waits implied by recorded dependencies. Sleep-induced intervals with no active target core additionally depend on wake-up timing, which is not determined by instruction costs or dependency order alone. The current model omits this component when accurate target sleep durations are unavailable. For eight-core x86 targets with one host worker, total-cycle MAPE relative to gem5 is 14.97\%, compared with 9.99\% for running cycles. The corresponding values are 22.24\% and 9.33\% for \emph{Alt.\ memory}, and 19.98\% and 6.73\% for \emph{Alt.\ core \& memory}. Incorporating these intervals requires wake-up timing consistent with the modeled target.

\section{Conclusion}
\label{sec:conclusion}
Long simulation times restrict the scale of architectural studies, especially when available host parallelism exceeds the target core count. \simname{} addresses this bottleneck by combining instruction-driven timing with a temporal design that uses synchronization information both to support independent replay and to reconstruct execution time. This decouples host replay order from target execution order, exposing additional parallelism while accounting for synchronization waits.

On a single host node, 96 workers deliver 1{,}010 MIPS, with a peak of 1{,}421 MIPS. The corresponding mean absolute percentage error in execution cycles relative to gem5 is 7.43\%. These results demonstrate the potential to shorten turnaround for repeated architectural timing analysis.

Future work will explore an execution-driven implementation of \simname{}, combining functional execution with independent local timing and dependency-based reconstruction of overall execution time. We will also investigate modeling additional synchronization mechanisms to broaden workload coverage. Moreover, we will explore methods for estimating suspension time to improve overall timing predictions.
\begin{acks}
OpenAI ChatGPT and Codex assisted with drafting and revising portions of the manuscript. The authors take full responsibility for the accuracy, originality, attribution, and integrity of the final content.
\end{acks}

\bibliographystyle{ACM-Reference-Format}
\bibliography{references}

@article{gem5,
  author  = {Binkert, Nathan and Beckmann, Bradford and Black, Gabriel
             and Reinhardt, Steven K. and Saidi, Ali and Basu, Arkaprava
             and Hestness, Joel and Hower, Derek R. and Krishna, Tushar
             and Sardashti, Somayeh and Sen, Rathijit and Sewell, Korey
             and Shoaib, Muhammad and Vaish, Nilay and Hill, Mark D.
             and Wood, David A.},
  title   = {The {gem5} Simulator},
  journal = {ACM SIGARCH Computer Architecture News},
  volume  = {39},
  number  = {2},
  pages   = {1--7},
  year    = {2011},
  doi     = {10.1145/2024716.2024718},
  url = {https://old.gem5.org/Publications.html}
}

@inproceedings{zsim,
  author    = {Sanchez, Daniel and Kozyrakis, Christos},
  title     = {{ZSim}: Fast and Accurate Microarchitectural Simulation
               of Thousand-Core Systems},
  booktitle = {Proceedings of the 40th Annual International Symposium
               on Computer Architecture},
  series    = {ISCA '13},
  pages     = {475--486},
  year      = {2013},
  publisher = {Association for Computing Machinery},
  doi       = {10.1145/2485922.2485963},
  url = {https://people.csail.mit.edu/sanchez/papers/2013.zsim.isca.pdf}
}

@inproceedings{interval,
  author = {Davy Genbrugge and Stijn Eyerman and Lieven Eeckhout},
  title = {Interval Simulation: Raising the Level of Abstraction in
    Architectural Simulation},
  booktitle = {IEEE International Symposium on High Performance Computer
    Architecture (HPCA)},
  year = {2010},
  url = {https://biblio.ugent.be/publication/845663}
}

@inproceedings{sniper,
  author    = {Carlson, Trevor E. and Heirman, Wim and Eeckhout, Lieven},
  title     = {{Sniper}: Exploring the Level of Abstraction for Scalable
               and Accurate Parallel Multi-Core Simulation},
  booktitle = {Proceedings of the International Conference for High
               Performance Computing, Networking, Storage and Analysis},
  series    = {SC '11},
  pages     = {52:1--52:12},
  year      = {2011},
  publisher = {Association for Computing Machinery},
  doi       = {10.1145/2063384.2063454},
  url = {https://snipersim.org/w/Paper:Sc2011Carlson}
}

@article{prophet,
  author = {Weihua Zhang and Xiaofeng Ji and Yunping Lu and Haojun Wang and
    Haibo Chen and Pen-Chung Yew},
  title = {{Prophet}: A Parallel Instruction-Oriented Many-Core Simulator},
  journal = {IEEE Transactions on Parallel and Distributed Systems},
  volume = {28},
  number = {10},
  pages = {2939--2952},
  year = {2017},
  doi = {10.1109/TPDS.2017.2700307},
  url = {https://experts.umn.edu/en/publications/prophet-a-parallel-instruction-oriented-many-core-simulator/}
}

@inproceedings{graphite,
  author    = {Miller, Jason E. and Kasture, Harshad and Kurian, George
               and Gruenwald, III, Charles and Beckmann, Nathan
               and Celio, Christopher and Eastep, Jonathan
               and Agarwal, Anant},
  title     = {{Graphite}: A Distributed Parallel Simulator for Multicores},
  booktitle = {Proceedings of the 16th IEEE International Symposium
               on High-Performance Computer Architecture},
  series    = {HPCA '10},
  pages     = {1--12},
  year      = {2010},
  publisher = {IEEE},
  doi       = {10.1109/HPCA.2010.5416635},
  url = {https://people.csail.mit.edu/beckmann/publications/papers/2010.hpca.graphite.pdf}
}

@article{synchrotrace,
  author = {Karthik Sangaiah and Michael Lui and Radhika Jagtap and
    Stephan Diestelhorst and Siddharth Nilakantan and Ankit More and
    Baris Taskin and Mark Hempstead},
  title = {{SynchroTrace}: Synchronization-Aware Architecture-Agnostic Traces
    for Lightweight Multicore Simulation of {CMP} and {HPC} Workloads},
  journal = {ACM Transactions on Architecture and Code Optimization},
  volume = {15},
  number = {1},
  pages = {2:1--2:25},
  year = {2018},
  doi = {10.1145/3158642},
  url = {https://research.coe.drexel.edu/ece/vlsi/images/b/b1/ST_2018.pdf}
}

@inproceedings{elastic,
  author = {Radhika Jagtap and Stephan Diestelhorst and Andreas Hansson and
    Matthias Jung and Norbert Wehn},
  title = {Exploring System Performance Using Elastic Traces:
    Fast, Accurate and Portable},
  booktitle = {International Conference on Embedded Computer Systems:
    Architectures, Modeling and Simulation (SAMOS)},
  year = {2016},
  url = {https://www.gem5.org/documentation/general_docs/cpu_models/TraceCPU}
}

@article{elasticsimmate,
  author = {Alejandro Nocua and Florent Bruguier and Gilles Sassatelli and
    Abdoulaye Gamati{\'e}},
  title = {A gem5 trace-driven simulator for fast architecture exploration
    of {OpenMP} workloads},
  journal = {Microprocessors and Microsystems},
  volume = {67},
  pages = {42--55},
  year = {2019},
  doi = {10.1016/j.micpro.2019.01.008},
  url = {https://doi.org/10.1016/j.micpro.2019.01.008}
}

@inproceedings{simpoint,
  author = {Timothy Sherwood and Erez Perelman and Greg Hamerly and Brad Calder},
  title = {Automatically Characterizing Large Scale Program Behavior},
  booktitle = {Proceedings of the Tenth International Conference on
    Architectural Support for Programming Languages and Operating Systems
    (ASPLOS)},
  year = {2002},
  url = {https://cseweb.ucsd.edu/~calder/abstracts/ASPLOS-02-SimPoint.html}
}

@inproceedings{qemu,
  author = {Fabrice Bellard},
  title = {{QEMU}, a Fast and Portable Dynamic Translator},
  booktitle = {2005 USENIX Annual Technical Conference, FREENIX Track},
  year = {2005},
  publisher = {USENIX Association},
  url = {https://www.usenix.org/conference/2005-usenix-annual-technical-conference/qemu-fast-and-portable-dynamic-translator}
}

@misc{spec2017,
  author = {{Standard Performance Evaluation Corporation}},
  title = {{SPEC CPU 2017}},
  howpublished = {\url{https://www.spec.org/cpu2017/}},
  note = {Accessed 23 October 2026}
}

@misc{parsec,
  author = {{Princeton University}},
  title = {The {PARSEC} Benchmark Suite},
  howpublished = {\url{https://parsec.cs.princeton.edu/}},
  note = {Project website}
}

@misc{npb,
  author = {{NASA Advanced Supercomputing Division}},
  title = {{NAS Parallel Benchmarks}},
  howpublished = {\url{https://www.nas.nasa.gov/software/npb.html}},
  note = {Accessed 23 October 2026}
}

@article{partigem5,
  author = {Jos{\'e} Cubero-Cascante and Niko Zurstra{\ss}en and
    J{\"o}rn N{\"o}ller and Rainer Leupers and Jan Moritz Joseph},
  title = {Scalable and Accurate Parallel Timing Simulations with {parti-gem5}},
  journal = {International Journal of Parallel Programming},
  volume = {54},
  note = {Article 16},
  year = {2026},
  doi = {10.1007/s10766-026-00819-5},
  url = {https://doi.org/10.1007/s10766-026-00819-5}
}

@inproceedings{smarts,
  author = {Roland E. Wunderlich and Thomas F. Wenisch and Babak Falsafi and James C. Hoe},
  title = {{SMARTS}: Accelerating Microarchitecture Simulation via Rigorous Statistical Sampling},
  booktitle = {International Symposium on Computer Architecture (ISCA)},
  year = {2003},
  url = {https://users.ece.cmu.edu/~jhoe/doku/doku.php?id=smarts_simulation_sampling}
}

@inproceedings{barrier-interval,
  author = {Paul D. Bryan and Jason A. Poovey and Jesse G. Beu and Thomas M. Conte},
  title = {Accelerating Multi-threaded Application Simulation through Barrier-Interval Time-Parallelism},
  booktitle = {IEEE International Symposium on Modeling, Analysis and Simulation of Computer and Telecommunication Systems (MASCOTS)},
  year = {2012},
  doi = {10.1109/MASCOTS.2012.23}
}

@inproceedings{barrierpoint,
  author = {Trevor E. Carlson and Wim Heirman and Kenzo Van Craeynest and Lieven Eeckhout},
  title = {{BarrierPoint}: Sampled Simulation of Multi-Threaded Applications},
  booktitle = {IEEE International Symposium on Performance Analysis of Systems and Software (ISPASS)},
  year = {2014},
  doi = {10.1109/ISPASS.2014.6844456}
}

@inproceedings{taskpoint,
  author = {Thomas Grass and Alejandro Rico and Marc Casas and Miquel Moret{\'o} and Eduard Ayguad{\'e}},
  title = {{TaskPoint}: Sampled Simulation of Task-Based Programs},
  booktitle = {IEEE International Symposium on Performance Analysis of Systems and Software (ISPASS)},
  year = {2016},
  doi = {10.1109/ISPASS.2016.7482104}
}

@inproceedings{looppoint,
  author = {Alen Sabu and Harish Patil and Wim Heirman and Trevor E. Carlson},
  title = {{LoopPoint}: Checkpoint-driven Sampled Simulation for Multi-threaded Applications},
  booktitle = {IEEE International Symposium on High-Performance Computer Architecture (HPCA)},
  year = {2022},
  doi = {10.1109/HPCA53966.2022.00051}
}

@article{looppoint-extension,
  author = {Alen Sabu and Zhantong Qiu and Harish Patil and Changxi Liu and Wim Heirman and Jason Lowe-Power and Trevor E. Carlson},
  title = {Accelerating the Simulation of Parallel Workloads using Loop-Bounded Checkpoints},
  journal = {ACM Transactions on Architecture and Code Optimization},
  volume = {23},
  number = {1},
  year = {2026},
  doi = {10.1145/3799430}
}

@inproceedings{mtr,
  author = {Kenneth C. Barr and Heidi Pan and Michael Zhang and Krste Asanovi{\'c}},
  title = {Accelerating Multiprocessor Simulation with a Memory Timestamp Record},
  booktitle = {IEEE International Symposium on Performance Analysis of Systems and Software (ISPASS)},
  year = {2005},
  url = {https://publications.csail.mit.edu/abstracts/abstracts05/kbarr/kbarr.html}
}

@article{iw-centric,
  author    = {Carlson, Trevor E. and Heirman, Wim and Eyerman, Stijn
               and Hur, Ibrahim and Eeckhout, Lieven},
  title     = {An Evaluation of High-Level Mechanistic Core Models},
  journal   = {ACM Transactions on Architecture and Code Optimization},
  volume    = {11},
  number    = {3},
  pages     = {28:1--28:25},
  year      = {2014},
  doi       = {10.1145/2629677}
}

@book{data-networks,
  author = {Dimitri P. Bertsekas and Robert G. Gallager},
  title = {Data Networks},
  edition = {2},
  publisher = {Prentice Hall},
  year = {1992},
  url = {https://web.mit.edu/dimitrib/www/datanets.html}
}

@inproceedings{rico2011trace,
  author    = {Rico, Alejandro and Duran, Alejandro and Cabarcas, Felipe
               and Etsion, Yoav and Ramirez, Alex and Valero, Mateo},
  title     = {Trace-Driven Simulation of Multithreaded Applications},
  booktitle = {Proceedings of the IEEE International Symposium
               on Performance Analysis of Systems and Software},
  series    = {ISPASS '11},
  pages     = {87--96},
  year      = {2011},
  publisher = {IEEE},
  doi       = {10.1109/ISPASS.2011.5762718}
}

@inproceedings{traceparallel1997,
  author    = {Nguyen, A.-T. and Bose, Pradip and Ekanadham, K.
               and Nanda, A. K. and Michael, M. M.},
  title     = {Accuracy and Speed-Up of Parallel Trace-Driven
               Architectural Simulation},
  booktitle = {Proceedings of the 11th International Parallel
               Processing Symposium},
  pages     = {39--44},
  year      = {1997},
  publisher = {IEEE},
  doi       = {10.1109/IPPS.1997.580842}
}

@article{chunkedtejas,
  author  = {Kalayappan, Rajshekar and Chhabra, Avantika
             and Sarangi, Smruti R.},
  title   = {{ChunkedTejas}: A Chunking-Based Approach to Parallelizing
             a Trace-Driven Architectural Simulator},
  journal = {ACM Transactions on Modeling and Computer Simulation},
  volume  = {30},
  number  = {3},
  pages   = {15:1--15:21},
  year    = {2020},
  doi     = {10.1145/3375397}
}

@misc{posix2018,
  author       = {{The Open Group}},
  title        = {The {Open Group Base Specifications}, Issue 7,
                  2018 Edition},
  year         = {2018},
  note         = {IEEE Std 1003.1-2017},
  url          = {https://pubs.opengroup.org/onlinepubs/9699919799/}
}

@misc{riscv-a-extension,
  author       = {{RISC-V International}},
  title        = {{RISC-V} Instruction Set Manual:
                  {A} Extension for Atomic Instructions,
                  Version 2.1},
  year         = {2024},
  note         = {Ratified specification},
  url          = {https://docs.riscv.org/reference/isa/v20240411/unpriv/a-st-ext.html}
}

@manual{arm-armv8,
  author       = {{Arm Limited}},
  title        = {{Arm Architecture Reference Manual for
                  A-profile Architecture}},
  number       = {DDI 0487},
  organization = {Arm Limited},
  url          = {https://support.arm.com/documentation/ddi0487/maa/}
}

@manual{intel-sdm,
  author       = {{Intel Corporation}},
  title        = {{Intel 64 and IA-32 Architectures
                  Software Developer's Manual}},
  organization = {Intel Corporation},
  url          = {https://www.intel.com/content/www/us/en/developer/articles/technical/intel-sdm.html}
}

@inproceedings{gem5-20,
  author    = {Lowe-Power, Jason and Ahmad, Abdul Mutaal and Akram, Ayaz
               and Alian, Mohammad and Amslinger, Rico and Andreozzi, Matteo
               and Armejach, Adri{\`a} and Asmussen, Nils and Beckmann, Bradford
               and Bharadwaj, Srikant and Black, Gabe and Bruce, Bobby R.
               and Carvalho, Daniel Rodrigues and Castrillon, Jeronimo
               and Chen, Lizhong and Derumigny, Nicolas and Diestelhorst, Stephan
               and Elsasser, Wendy and Fariborz, Marjan and Farmahini-Farahani, Amin
               and Fotouhi, Pouya and Gambord, Ryan and Gandhi, Jayneel
               and Grass, Thomas and Hansson, Andreas and Haria, Swapnil
               and Hayes, Timothy and Jafri, Syed Ali Raza and Jones, Timothy M.
               and Jung, Matthias and Kannoth, Subash and Krishna, Tushar
               and Lowe-Power, Jason and Menard, Christian and Moreto, Miquel
               and Nikoleris, Nikos and Olson, Lena E. and Orr, Marc
               and Wood, David A.},
  title     = {The {gem5} Simulator: Version 20.0+},
  year      = {2020},
  eprint    = {2007.03152},
  archivePrefix = {arXiv},
  primaryClass = {cs.AR},
  url       = {https://arxiv.org/abs/2007.03152}
}

@article{akram2019survey,
  author  = {Akram, Ayaz and Sawalha, Lina},
  title   = {A Survey of Computer Architecture Simulation Techniques and Tools},
  journal = {IEEE Access},
  volume  = {7},
  pages   = {78120--78145},
  year    = {2019},
  doi     = {10.1109/ACCESS.2019.2917698}
}

@inproceedings{shao2016co,
  author    = {Shao, Yakun Sophia and Xi, Sam Likun
               and Srinivasan, Vijayalakshmi
               and Wei, Gu-Yeon and Brooks, David M.},
  title     = {Co-Designing Accelerators and {SoC} Interfaces
               Using {gem5-Aladdin}},
  booktitle = {Proceedings of the 49th Annual IEEE/ACM
               International Symposium on Microarchitecture},
  series    = {MICRO '16},
  pages     = {48:1--48:12},
  year      = {2016},
  publisher = {IEEE},
  doi       = {10.1109/MICRO.2016.7783751}
}

@inproceedings{huppert2021memory,
  author    = {Huppert, Quentin and Evenblij, Timon
               and Perumkunnil, Manu and Catthoor, Francky
               and Torres, Lionel and Novo, David},
  title     = {Memory Hierarchy Calibration Based on Real Hardware
               In-Order Cores for Accurate Simulation},
  booktitle = {Proceedings of the Design, Automation and Test
               in Europe Conference and Exhibition},
  series    = {DATE '21},
  pages     = {707--710},
  year      = {2021},
  publisher = {IEEE},
  doi       = {10.23919/DATE51398.2021.9474108}
}

@inproceedings{mcqemu,
  author    = {Carvalho, Humberto and Nelissen, Geoffrey
               and Zaykov, Pavel},
  title     = {mc{QEMU}: Time-Accurate Simulation of Multi-core
               Platforms Using {QEMU}},
  booktitle = {Proceedings of the 23rd Euromicro Conference
               on Digital System Design},
  series    = {DSD '20},
  pages     = {81--88},
  year      = {2020},
  publisher = {IEEE},
  doi       = {10.1109/DSD51259.2020.00024}
}

@book{hennessy2017computer,
  author    = {Hennessy, John L. and Patterson, David A.},
  title     = {Computer Architecture: A Quantitative Approach},
  edition   = {6},
  publisher = {Morgan Kaufmann},
  year      = {2017},
  isbn      = {9780128119051}
}

@inproceedings{girbal2003dist,
  author    = {Girbal, Sylvain and Mouchard, Gilles and Cohen, Albert and Temam, Olivier},
  title     = {{DiST}: A Simple, Reliable and Scalable Method to Significantly Reduce Processor Architecture Simulation Time},
  booktitle = {Proceedings of the 2003 ACM SIGMETRICS International Conference on Measurement and Modeling of Computer Systems},
  year      = {2003},
  publisher = {Association for Computing Machinery},
  doi       = {10.1145/781028.781029}
}

@inproceedings{perelman2003simpoint,
  author    = {Perelman, Erez and Hamerly, Greg and
               Van Biesbrouck, Michael and Sherwood, Timothy and
               Calder, Brad},
  title     = {Using {SimPoint} for Accurate and Efficient Simulation},
  booktitle = {Proceedings of the 2003 ACM SIGMETRICS International
               Conference on Measurement and Modeling of Computer Systems},
  pages     = {318--319},
  year      = {2003},
  doi       = {10.1145/781027.781076}
}

@article{haskins2005warmup,
  author  = {Haskins, Jr., John W. and Skadron, Kevin},
  title   = {Accelerated Warmup for Sampled Microarchitecture Simulation},
  journal = {ACM Transactions on Architecture and Code Optimization},
  volume  = {2},
  number  = {1},
  pages   = {78--108},
  year    = {2005},
  doi     = {10.1145/1061267.1061272}
}

@inproceedings{calipers,
  author    = {Golestani, Hossein and Sen, Rathijit and Young, Vinson and Gupta, Gagan},
  title     = {{Calipers}: A Criticality-aware Framework for Modeling Processor Performance},
  booktitle = {Proceedings of the 36th ACM International Conference on Supercomputing},
  year      = {2022},
  pages     = {1--14},
  doi       = {10.1145/3524059.3532390}
}

@article{splitsim,
  author  = {Li, Hejing and Meiers, Marvin and Li, Jialin and Kaufmann, Antoine},
  title   = {{SplitSim}: Towards Practical Large-Scale Full-System Simulation for Systems Research},
  journal = {Proceedings of the ACM on Networking},
  volume  = {3},
  number  = {CoNEXT4},
  pages   = {52:1--52:19},
  year    = {2025},
  doi     = {10.1145/3768999}
}

@inproceedings{legosim,
  author    = {Lin, Tiantian and Qiu, Cheng and Wang, Xiaohang and Wang, Ling and Zheng, Zhulin and Jiang, Yingtao and Singh, Amit Kumar and Yin, Jieming and Qiu, Sihai and Li, Xiaodong and Tang, Xin and Song, Jie and Zhang, Mingzhe and Ren, Kui},
  title     = {{LEGOSim}: A Unified Parallel Simulation Framework for Multi-chiplet Heterogeneous Integration},
  booktitle = {Proceedings of the 58th IEEE/ACM International Symposium on Microarchitecture},
  year      = {2025},
  pages     = {1347--1362},
  doi       = {10.1145/3725843.3756068}
}

@inproceedings{pandey2022gpu,
  author    = {Pandey, Santosh and Li, Lingda and Flynn, Thomas and Hoisie, Adolfy and Liu, Hang},
  title     = {Scalable Deep Learning-Based Microarchitecture Simulation on {GPUs}},
  booktitle = {SC22: International Conference for High Performance Computing, Networking, Storage and Analysis},
  year      = {2022},
  pages     = {1--15},
  doi       = {10.1109/SC41404.2022.00084}
}

@inproceedings{hetsim,
  author    = {Pal, Subhankar and Kaszyk, Kuba and Feng, Siying and Franke, Bj{\"o}rn and Cole, Murray and O'Boyle, Michael and Mudge, Trevor and Dreslinski, Ronald G.},
  title     = {{HETSIM}: Simulating Large-Scale Heterogeneous Systems using a Trace-driven, Synchronization and Dependency-Aware Framework},
  booktitle = {2020 IEEE International Symposium on Workload Characterization},
  year      = {2020},
  pages     = {13--24},
  doi       = {10.1109/IISWC50251.2020.00011}
}

@article{simnet,
  author  = {Li, Lingda and Pandey, Santosh and Flynn, Thomas and Liu, Hang and Wheeler, Noel and Hoisie, Adolfy},
  title   = {{SimNet}: Accurate and High-Performance Computer Architecture Simulation using Deep Learning},
  journal = {Proceedings of the ACM on Measurement and Analysis of Computing Systems},
  volume  = {6},
  number  = {2},
  pages   = {25:1--25:24},
  year    = {2022},
  doi     = {10.1145/3530891}
}

@article{tao,
  author  = {Pandey, Santosh and Yazdanbakhsh, Amir and Liu, Hang},
  title   = {{TAO}: Re-Thinking {DL}-based Microarchitecture Simulation},
  journal = {Proceedings of the ACM on Measurement and Analysis of Computing Systems},
  volume  = {8},
  number  = {2},
  pages   = {28:1--28:25},
  year    = {2024},
  doi     = {10.1145/3656012}
}

@inproceedings{concorde,
  author    = {Nasr-Esfahany, Arash and Alizadeh, Mohammad and Lee, Victor and Alam, Hanna and Coon, Brett W. and Culler, David and Dadu, Vidushi and Dixon, Martin and Levy, Henry M. and Pandey, Santosh and Ranganathan, Parthasarathy and Yazdanbakhsh, Amir},
  title     = {{Concorde}: Fast and Accurate {CPU} Performance Modeling with Compositional Analytical-{ML} Fusion},
  booktitle = {Proceedings of the 52nd Annual International Symposium on Computer Architecture},
  pages     = {1480--1494},
  year      = {2025},
  doi       = {10.1145/3695053.3731037}
}

@inproceedings{musa,
  author    = {Grass, Thomas and Allande, C{\'e}sar and Armejach, Adri{\`a} and Rico, Alejandro and Ayguad{\'e}, Eduard and Labarta, Jes{\'u}s and Valero, Mateo and Casas, Marc and Moret{\'o}, Miquel},
  title     = {{MUSA}: A Multi-Level Simulation Approach for Next-Generation {HPC} Machines},
  booktitle = {SC16: International Conference for High Performance Computing, Networking, Storage and Analysis},
  pages     = {526--537},
  year      = {2016},
  doi       = {10.1109/SC.2016.44}
}

@inproceedings{chronos,
  author    = {Lin, Shanqing and Li, Yuanlong and Alian, Mohammad and Falsafi, Babak and Lai, Konrad K.},
  title     = {Accurate \& Scalable Time-Based Sampling of Full-System Simulation},
  booktitle = {Proceedings of the 59th IEEE/ACM International Symposium on Microarchitecture},
  year      = {2026},
  url       = {https://infoscience.epfl.ch/handle/20.500.14299/267412}
}

@inproceedings{tracertl,
  author    = {Zhang, Zifei and Xu, Yinan and Wang, Sa and Tang, Dan and Bao, Yungang},
  title     = {{TraceRTL}: Agile Performance Evaluation for Microarchitecture Exploration},
  booktitle = {2026 IEEE International Symposium on High-Performance Computer Architecture (HPCA)},
  year      = {2026},
  doi       = {10.1109/HPCA68181.2026.11408626}
}

@misc{fcs,
  author       = {Lin, Tiantian and Long, Tengjun and Zhang, Mingzhe and Jiang, Yingtao and Wang, Xiaohang and Zhang, Chao and Singh, Amit Kumar and Yin, Jieming and Ren, Kui},
  title        = {{FCS}: Frontier-based Commit-time Synchronization Method for Modular Simulators},
  howpublished = {Author-posted abstract},
  year         = {2027},
  note         = {Listed as forthcoming at HPCA 2027},
  url          = {https://mingzhe-zhang.github.io/publications/HPCA2027-2}
}
\end{document}